\documentclass[10pt,prl,showpacs,twocolumn,superscriptaddress]{revtex4}
\usepackage{graphics,color,amsmath,amssymb,graphicx,amsfonts}
\usepackage{times}

\begin{document}

\title{Topological degeneracy and decoherence-protected qubits in circuit QED}

\author{Gang Chen}
\affiliation{Institute of Theoretical Physics, Shanxi University, Taiyuan 030006, China}
\affiliation{Department of Physics, Shaoxing College of Arts and Sciences, Shaoxing
312000, China}

\author{Zheng-Yuan Xue}
\affiliation{Department of Physics and Center of Theoretical and Computational
Physics,The University of Hong Kong, Pokfulam Road, Hong Kong, China}

\author{J.-Q. Liang} \email{jqliang@sxu.edu.cn}
\affiliation{Institute of Theoretical Physics, Shanxi University,
Taiyuan 030006, China}

\begin{abstract}
We introduce an extended Dicke model with controllable long-range
atom-atom interaction to simulate topologically ordered states and
achieve decoherence-protected qubits. We illustrate our idea in an
experimentally feasible circuit quantum electrodynamics scenario.
Due to the intrinsic competition with the atom-field coupling
strength, we first demonstrate that this atom-atom interaction can
exhibit a novel topological quantum interference effect arising from
the instanton and anti-instanton tunneling paths. As a consequence,
this proposed model only with a few odd-number of atoms has a
two-fold absolute degenerate ground-subspace with a large energy
gap, which can become larger with the increasing of the system-size.
It may also support the excitation of anyonic statistics, and thus
can be regarded as a possible candidate for processing topological
quantum memory.
\end{abstract}

\maketitle

Cooper-pair box in circuit quantum electrodynamic (QED) is a
promising scenario for physical implementation of quantum
computation \cite{Blais}. One of the advantages in such a device is
that the Cooper-pair box can be  acted, under certain conditions, as
an two-level artificial atom with well controlled parameters by
external currents, voltages and magnetic flux \cite{Makhlin, You1}.
Moreover, the strong coupling limit between Cooper-pair box and
cavity field has been experimentally achieved  \cite {Wallraff}. In
this implementation scenario, the cavity can provide strong
inhibition of spontaneous emission and suppress greatly the
decoherence caused by the external environment.

Since the larger-scale systems are more sensitive to decoherence,
the scaling of quantum information processors is a great challenge
to build a practical quantum computer. Therefore, it is a crucial
task to establish the decoherence-protected qubits, more preferably
in the physical level. Recently, a conceptually different approach
called topological quantum computing based on the braiding
operations of anyons seems to be a very promising strategy towards
fault-tolerance \cite{Kitaev}. The topological degenerate ground
states, resulted from the fractional  anyonic statistics and gapped
from excitations, is a basic condition to process topological
quantum memory \cite{Dennis}. Two-dimensional spin-lattice models
with nearest-neighbor coupling links in real space are fascinating
models to study the topological degeneracy and anyonic statistics
\cite{Kitaev}. The  model Hamiltonian can be simulated by ultracold
trapped atoms \cite{Duan} or polar molecules \cite{Micheli} in
optical lattices with highly coherent controllability and
measurement. Furthermore, anyonic interference in these systems has
been suggested to be observed by using local \cite{Zhang1} or global
operations \cite{Jiang}. On the other hand, the degenerate ground
states of quantum dimer-liquid realized in the Cooper-pair box
arrays \cite{Ioffe,Doucot} and of two dimensional Hamiltonian of
long-range coupling generated by trapped ions \cite{Milman} as well
as Cooper-pair boxes \cite{Xue} can be also served as topologically
decoherence-protected qubits. Generally, the degeneracy can be
removed by quantum tunneling, which results in a tunnel splitting
proportional to $\exp [-N]$ with $N$ being the number of atoms
\cite{Leggett}. It is clear that the tunnel splitting vanishes in
the large $N$ limit. However, it is very difficult to fabricate or
control large indistinguishable natural or artificial atoms with
current experimental technique.

In this paper, we introduce a new Hamiltonian, in which we term as
an extended Dicke model, with controllable long-range atom-atom
interaction to simulate topologically ordered states and achieve
decoherence-protected qubits. The motivation for our consideration
is given as follows. (1) The topological degeneracy is absolute for
odd-number of atoms $(N=3,5,...)$ with vanishing tunnel splitting.
This is resulted from a novel topological quantum interference of
the instantons and anti-instanton tunneling paths, attributed to the
intrinsic competition between the atom-atom interaction and the\
atom-field coupling strength. (2) A large energy gap can be
generated and becomes larger with the increasing of the system-size,
which is different from the spin-lattice models and very useful in
designing the topologically protected qubits for quantum computing.
(3) This Hamiltonian supports the excitation of anyonic statistics
and the braiding operations can be constructed by the trajectories
of instantons on an effective circular-space. (4) This Hamiltonian
can be realized in practical superconducting circuit with only one
step. The superconducting  quantum network with four artificial
atoms was experimentally achieved \cite{Grajcar}.

\begin{figure}[tbh]
\includegraphics[width=8cm]{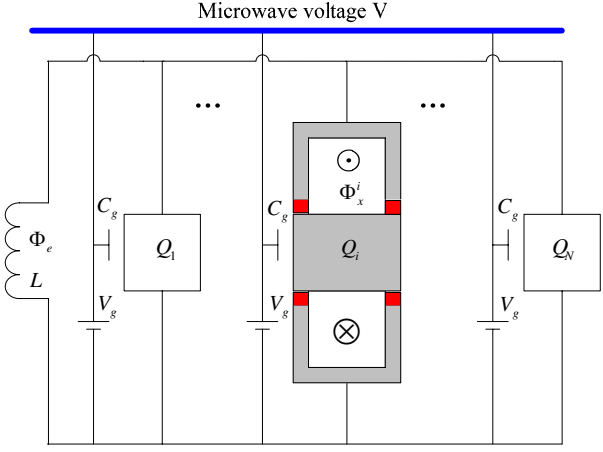}\newline
\caption{(Color online) Schematic diagram of an experimental setup,
where identical Cooper-pair boxes weakly coupled by two symmetric
superconducting quantum interference devices are connected in
parallel to a common superconducting inductance. Each Josephson
junction (Red area) is operated both in the charging regime and at
low temperature. The inductance is here chosen to be so large that
the inductance of the circuit except for the inductance can be
neglected, if the circuit is not too large. The one-dimensional
transmission line cavity (Blue line) can generate a microwave
voltage, which has been successfully achieved in experiment.}
\label{fig1}
\end{figure}

We first consider the implementation of our model Hamiltonian with
superconducting circuit as shown in Fig. \ref{fig1}. Cooper-pair
box, consisting of two superconducting quantum interference devices
(SQUIDs), is acted as a two-level artificial atom.  All the
identical atoms are biased by both a dc voltage $V_{g}$ and a
microwave voltage $V$ through the same gate capacitance $C_{g}$. The
Hamiltonian for free atoms can be
written as ($\hbar =1$ hereafter) \cite{Makhlin, You1}: $H_{q}=\sum_{i=1}^{N}%
\varepsilon \sigma _{z}^{i}+\eta \sigma _{x}^{i}\ $with $\varepsilon
=2E_{c}(2n_{g}-1)$ and $\eta =-E_{J}\cos \Phi _{x}\cos \Phi
_{e}(1-2\varkappa ^{2}\sin ^{2}\Phi _{e})$, where $E_{c}$ is the
charging energy, $E_{J}$ is the Josephson energy,
$n_{g}=C_{g}V_{g}/2e$ is the induced charge, $\Phi _{x}$ is the
local magnetic flux through the two SQUID loops in each Cooper-pair
box and is assumed to have the same values but opposite directions,
$\Phi _{e}$ is the global magnetic flux through the common
superconducting inductance $L$, and $\varkappa =\pi LI_{c}/\Phi
_{0}$ with $I_{c}$ being the critical current and $\Phi _{0}$ being
the flux quantum. The atoms are capacitively coupled to the
single-mode cavity field of frequency $\omega $ at the voltage
maximum, yielding a strong electric dipole interaction \cite{Blais}
such that $H_{qc}=\sum_{i=1}^{N}g(a^{\dagger }+a)(\sigma
_{z}^{i}+2n_{g}-1)$, where $a$ ($a^{\dagger }$) is the annihilation
(creation) operator for the cavity mode and $g$ is the atom-field
coupling strength. The phase changes in the SQUID-loops can be
written as $\varphi _{k}=(\varphi _{k}^{l}+\varphi _{k}^{r})/2$ with
$k=d,u$ representing the down and up loops and $l,r$ denoting the
left and right junctions in each SQUID-loop. The phase changes of
the left junctions $\varphi _{u}^{l}$ and $\varphi _{d}^{l}$ are
related to the total flux $\Phi =\Phi _{e}+LI$ through the
inductance $L$ by the constraint: $\varphi _{d}^{L}-\varphi
_{u}^{L}=2\pi \Phi /\Phi _{0}\equiv 2\phi $, where $I=\sum_{i}I_{i}$
is the total current through the inductance $L$. Since the local
magnetic fluxes $\Phi _{x}$ are assumed to have the same values but
opposite directions, the phases induced by the flux-pair cancel each
other in any loop embracing both of them leading to a constraint
$\varphi _{d}-\varphi _{u}=2\phi $. Therefore, the two-body
interaction of atoms comes from the inductive coupling \cite{You2}:
$H_{qq}=\frac{1}{2}LI^{2}\approx -2v\sum_{j>i=1}^{N}\sigma
_{x}^{i}\sigma _{x}^{j}$ with $v=(LI_{c}^{2}\sin ^{2}\Phi _{e})/2$,
which shows that the long-range atom-atom interaction is well
controlled by the global magnetic flux $\Phi _{e}$. By using the
collective spin operator $S_{\mu}=\sum_{i=1}^{N}\sigma_{\mu}^{i}$
with ${\mu}\in\{x, y, z\}$ and the total spin quantum number
$S=N/2$, considering the optimal point $(n_{g}=1/2)$ and $\cos
\Phi_{x}=0$, then the total Hamiltonian
$H=H_{c}+H_{qc}+H_{q}+H_{qq}$ with $H_{c}=\omega a^{\dagger }a$ can
be simplified as
\begin{equation}
H=\omega a^{\dagger }a+g(a^{\dagger }+a)S_{z}-vS_{x}^{2}.  \label{1}
\end{equation}%
Hamiltonian (\ref{1}) is our extended Dicke model with the
long-range atom-atom interaction (the last term) and can be here
suggested to simulate topologically ordered states and achieve
decoherence-protected qubits. For typical experimental parameters
with the frequency $\omega =2\pi \times 6\times 10^{3}$ MHz and the
atom-field coupling strength $g=2\pi \times 5.8$ MHz
\cite{Wallraff}, the condition $u=g^{2}/\omega<v$, which plays a
crucial role in the following calculations, can be well satisfied
since the maximum atom-atom interaction constant can be achieved at
$v=10^{2}$ MHz for large inductance $L=10$ nH \cite{Chen}.

We then demonstrate that Hamiltonian (1) has an absolute topological
degeneracy for odd-number of atoms and a large energy gap by means
of the coherent-state path integral \cite{Zhang}. In the
coherent-state representation for both the field $\left\vert \alpha
\right\rangle =\left\vert x+iy\right\rangle$ and the atoms
$\left\vert \Omega \right\rangle =\left\vert \theta, \varphi
\right\rangle$, the semiclassical energy of the system can be
obtained by
\begin{equation}
E=\omega (x^{2}+y^{2})+2Sgx\cos \theta -vS^{2}\sin ^{2}\theta \cos
^{2}\varphi .  \label{2}
\end{equation}%
For $u<v$, it is obvious that the semiclassical energy in Eq.
(\ref{2}) has two-fold degenerate ground-states corresponding to the
giant-spin orientations $\theta =\pi /2,\varphi =0$ and $\theta =\pi
/2,\varphi =\pi $, at the vacuum state of the cavity-field with
$x=y=0$. We may denote the two degenerate orientations of the
giant-spin, i.e., macroscopic quantum states, as $\left\vert 0
\right\rangle$ and $\left\vert \pi \right\rangle$, respectively.

It is well known that quantum tunneling can occur between the
degenerate vacua $\left\vert 0\right\rangle$ and $\left\vert \pi
\right\rangle$. As a consequence, the degeneracy is broken and two
low-lying eigenstates with a small tunnel splitting $d$ are formed
by symmetric and antisymmetric superpositions of the macroscopic
quantum states (called the Schr\"{o}dinger cat states) such that
$\left\vert \psi _{\pm }\right\rangle =(\left\vert 0\right\rangle
\pm \left\vert \pi \right\rangle )/\sqrt{2}$ \cite{Leggett}. We here
reexamine the tunnel splitting $d$, which is inversely proportional
to the imaginary-time transition-amplitude \cite{Zhang}
\begin{eqnarray}
P &=&\left\langle \pi ,\beta \right\vert \exp (-\beta H)\left\vert
0,0\right\rangle  \notag \\
&=&\int \int \mathcal{D}\{\alpha \}\mathcal{D}\{\Omega \}\exp (-S_{E}),
\label{3}
\end{eqnarray}%
where $\beta $ is the imaginary time-period, and $S_{E}=\lim_{\beta
\rightarrow \infty }\int_{0}^{\beta }\mathcal{L}_{E}d\tau $ with $\ \tau =it$
is the Euclidean action evaluated along the tunneling trajectory of
instanton, which, in the context of quantum field theory, may be visualized
a pseudoparticle moving between degenerate vacua under the barrier region
and has nonzero topological charge but zero energy. The corresponding
Lagrangian is given by
\begin{equation}
\mathcal{L}_{E}=iS(1-\cos\theta)\dot{\varphi}+i(x\dot{y}-\dot{x}y)+E.
\label{4}
\end{equation}%
The first two parts in Eq. (\ref{4}) are seen to be the Wess-Zumino
interactions \cite{Fradkin}, which have a profound topological
origin giving rise to phase-interference between different tunneling
paths, and play a crucial role in the generation of
decoherence-protected qubits as shown in the following.

The first Wess-Zumino term in Eq. (\ref{4}) results in a topological
boundary condition
\begin{equation}
iS\int_{0}^{\beta }\dot{\varphi}d\tau =iS[\varphi (\beta )-\varphi (0)+2n\pi
],  \label{5}
\end{equation}%
where $n$ is winding number counting the number of times that the
path wraps around the north pole. The contribution of any path on
the Bloch sphere $S_{2}$, described by the parameters $\theta (\tau
)$ and $\varphi (\tau )$, to the Euclidean action $S_{E}$ is equal
to $iS$ times the area swept out between the path and the north
pole. For closed paths, this is identical to the Berry phase
\cite{Berry}.

\begin{figure}[ptb]
\includegraphics[width=8cm]{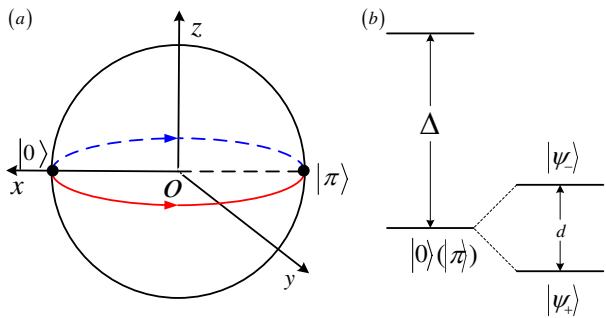}\newline
\caption{(Color online) (a)The interference of an instanton (Red
line) and an anti-instantons (Blue dotted line) tunneling paths from
$\left\vert 0\right\rangle $ to $\left\vert \protect\pi
\right\rangle $ in the surface of the Bloch sphere. (b) The energy
gap $\Delta $ and the tunnel splitting $d$ for odd- and even-
numbers of atoms, respectively.} \label{fig2}
\end{figure}

Hamiltonian (\ref{1}) possesses an important feature that the
passage from one energy-minimum $(\varphi =0)$ to the other
$(\varphi =\pi )$ can be accomplished either by an instanton or an
anti-instanton path [a clockwise or a counterclockwise tunneling
path on the Bloch sphere $S_{2}$ as shown in Fig. \ref{fig2} (a)]
and the transition amplitude $P$ is evaluated by a sum over paths
comprising a sequence of instantons and anti-instantons. Because of
symmetry, contribution of instantons and anti-instantons to $P$ are
the same except a phase-factor, which results in the quantum phase
interference \cite{Liang}, namely,
\begin{equation}
P=A\left\vert (e^{iS\pi }+e^{-iS\pi })\right\vert
e^{-S_{E}^{cl}}=2A\left\vert \cos (S\pi )\right\vert e^{-S_{E}^{cl}},
\label{6}
\end{equation}%
where $A$ is a constant and $S_{E}^{cl}$ proportional to $N$ is the action
along the classical trajectory of instanton.

\begin{figure}[ptb]
\includegraphics[width=8cm]{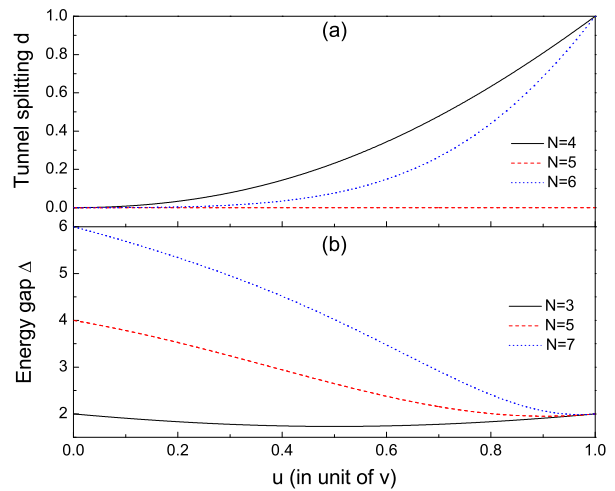}\newline
\caption{(Color online) The energy gap $\Delta $ (a) and the tunnel
splitting $d$ (b) in terms of the direct numerical simulation of
Hamiltonian (1)} \label{fig3}
\end{figure}

Interestingly, the transition amplitude $P$ in Eq. (\ref{6}) is
nonzero for even-number of atoms because of the $\cos (S\pi )$
factor, which arises directly from the topological boundary
condition in Eq. (\ref{5}) and the interference of the instanton and
anti-instanton tunneling paths (a generalized Aharnonov-Bohm
effect). More importantly, the original degeneracy of the states
$\left\vert 0\right\rangle \ $and $\left\vert \pi \right\rangle $ is
removed, and a tunnel splitting $d$ between the ground state levels
$\left\vert \psi _{\pm }\right\rangle $ exists. It means that in
this case the topological degeneracy is not absolute but precise for
the large $N$. However, when the number of the atoms is odd,
topological quantum interference is destructive, leading to a
vanishing transition amplitude since $\cos (S\pi )=0$. In other
words, the quantum tunneling between the states $\left\vert
0\right\rangle \ $and $\left\vert \pi \right\rangle $ is suppressed:
the tunnel splitting $d$ remains zero, which implies that the
degenerate ground states are stability for \textit{any} odd-number
of atoms. Fig. \ref{fig2}(b) shows a graphic illustration of the
energy gap $\Delta $ and the tunnel splitting $d$, which agree with
the numerical result of eigenenergies presented in Fig. \ref{fig3}.
It can be also shown in Fig. \ref{fig3} that for odd-number of atoms
the energy gap $\Delta $ becomes larger with the increasing of the
system-size but remains invariant when $u=v$, where Hamiltonian
(\ref{1}) is completely diagonalized. This novel property, in
contrast with the spin-lattice models that the energy gap turns into
smaller when the system size is increasing, is very useful in
designing the topologically protected qubits for quantum computing.
It is worth while to remark that the quantum tunneling and related
quantum phase interference are surely due to the long-range
atom-atom interaction. This phenomenon of
\textit{interaction-induced topological quantum interference} is
directly related to the Kramers degeneracy and the spin-parity
effect in magnetic systems \cite{Loss,Delft}.

We now illustrate in terms of the predicted particle-number parity
effect that the system only with odd-number of atoms supports the
excitation of anyonic statistics. From the topological boundary
condition in Eq.(5), the wavefunctions change a phase under $2\pi $
rotation, i.e.,
\begin{equation}
\psi (\varphi +2\pi )=\exp (2\pi iS)\psi (\varphi ),  \label{7}
\end{equation}%
which is periodic for even number of atoms and antiperiodic for
odd-number of atoms. In other words, the intanstan can be regarded
as a spinless fermion and boson for the odd- and even- numbers of
atoms, respectively. For the fermion case the paths from $\left\vert
0\right\rangle \ $to $\left\vert \pi \right\rangle $\ on equatorial
circle of the Bloch sphere belonging to different homotopy classes
$\xi $ cannot be continuously deformed into each other. Therefore,
the total transition amplitude $P$ can be reorganized a sum over
partial amplitudes $P_{\xi }$ of a distinct homotopy class with a
weight phase factor $\chi (\xi )$ characterizing the statistics
\cite{Laidlaw}:
\begin{equation}
P=\sum_{\xi \text{ }\epsilon \text{ }\pi _{1}}\chi (\xi )P_{\xi }
\label{8}
\end{equation}%
with $\pi _{1}$ being the first homotopy group formed by the
inequivalent paths. It should be noticed that these phase factors
$\chi (\xi )$ depending on homotopy classes actually form a
representation of the first homotopy group $\pi _{1}$ with the role
of multiplication $\chi (\xi _{1})\chi (\xi _{2})=\chi (\xi _{1}\xi
_{2})$, which leads to a solution $\chi (\xi )=\exp (\pm
i\tilde{\theta})$ with $0\leq \tilde{\theta}\leq \pi $. On the other
hand, the first homotopy group $\pi _{1}$ is identical to the braid group $%
B_{n}$ on $n$ strands in the two-dimensional configuration space
\cite{Wu}, which plays a central role in the anyonic statistics. If
we envision each fermion world line as a thread, each classical
trajectory contributing to the action $S_{E}^{cl}$ in Eq. (6)
becomes a \textit{braid}, where each fermion on the initial time
slice can be connected by a thread to any one of the fermion on the
final time slice. We may imagine that all fermions occupy
$n$ ordered positions (labeled by $1,2,3,...,n$) arranged on a line. Let $%
\eta _{1}$ denote a counterclockwise exchange of the two fermions
initially at positions $1$ and $2$, $\eta _{2}$ denote a
counterclockwise exchange of the two fermions initially at positions
$2$ and $3$, and so on. Any braid operation can be constructed as a
succession of exchanges of neighboring
fermions in terms of the group generators $\eta _{1}$, $\eta _{2}$, . . ., $%
\eta _{n-1}$, which satisfy two kind of relations \cite{Wu}:
\begin{equation}
\eta _{j}\eta _{k}=\eta _{k}\eta _{j},k\neq j\pm 1\text{; }\eta
_{j}\eta _{j+1}\eta _{j}=\eta _{j+1}\eta _{j}\eta _{j+1}.  \label{9}
\end{equation}

Finally, we explain how to use the system of $N$-two-level-atom in
the cavity field to design the topologically decoherence-protected
qubits based on the above solutions. In the temperature $T$ much
lower than the energy gap $\Delta $ the effect of environment is
negligible and the system evolves only within the ground state
subspace. Robustness against decoherence in short time process can
be achieved within the ground-state transition similar to the idea
used in adiabatic quantum computing \cite{Lidar}. Also,
decoherence-protected implementation of the basic quantum
logic-operation can be realized by the topological order
\cite{Schmidt-Kaler}.

In summary, we have introduced a new model with only a few atoms to
simulate the topologically ordered state and achieve
decoherence-protected qubits, which is attributed to the
interaction-induced topological quantum interference of the
instanton and anti-instanton tunneling paths. An experimentally
feasible scenario is proposed to realize the model Hamiltonian.
Since all artificial atoms work at their optimal point, this device
can be also mostly immune from charge noise produced by
uncontrollable charge fluctuations. Apart from designing the
required qubits, the anyonic interference may be also detected by
measuring just microwave photon statistics of the cavity.

We thank L. Jiang, L. B. Shao and H. B. Xue for helpful discussions
and suggestions. This work was supported by the Natural Science
Foundation of China under Grant Nos. 10775091 and 10704049.

\end{document}